\documentclass[12pt]{article} 
\usepackage{amssymb}
\usepackage{graphicx}
\newcommand{\preprintno}[1]
{\vspace{-2cm}{\normalsize\begin{flushright}#1\end{flushright}}\vspace{1cm}}

\title{\preprintno{{\bf SUSX-TH/00-009}}
The SQCD vacuum coupled to supergravity and string theory moduli}
\author{Thomas Dent\thanks{E-mail address: t.e.dent@sussex.ac.uk} \\ 
	{\em Centre for Theoretical Physics, University of Sussex,} \\
	{\em Brighton BN1 9QH, U.K.}}
\date{June 2000}

\begin{document} 
\maketitle 
\begin{abstract} 
\noindent 
We calculate the scalar potential of supersymmetric QCD (in the regime $N_f<N_c$) coupled to 
$\mathcal{N}=1$ supergravity with moduli-dependent gauge kinetic function and masses. The gauge dynamics 
are described by the Taylor-Veneziano-Yankielowicz superpotential for composite effective fields. The 
potential can be expanded about the ``truncated'' point in the gaugino and matter condensate directions in 
order to find corrections to the globally supersymmetric minimum. The results are relevant to the 
phenomenology of supersymmetry-breaking in string-inspired supergravity models, and also to recent work on 
domain walls in SQCD.
\end{abstract}
\section{Introduction}
Supersymmetric gauge theory with matter representations plays a central role in many models of
supersymmetry-breaking relevant to string phenomenology. In $\mathcal{N}=1$ supergravity models based on
perturbative string constructions, the gauge dynamics in a ``hidden sector'' \cite{DerenIN+DineRSW+Taylor85}
can induce a hierarchically small vacuum expectation value (v.e.v.)\ for the gaugino bilinear operator 
$\mbox{Tr}\,\lambda^{\alpha}\lambda_{\alpha}$ (the ``gaugino condensate''). If there are hidden matter 
multiplets \cite{LustT91,deCarlosCM91}, there may also be nonzero v.e.v.'s for the squark bilinears 
$\bar{\tilde{q}}_{ai} \tilde{q}^a_j$ where $a$ is the gauge group representation index and $i,j$ are 
flavour indices. When coupled to supergravity the gaugino condensate is a source of supersymmetry-breaking
\cite{Nilles+FerraraGN} which can in principle explain the ratio of the electroweak scale, and the 
notional masses of supersymmetric partners, to the mass scale of string theory. The scale of SUSY-breaking 
can be parameterized by the gravitino mass, $m_{3/2}$, which is proportional to 
$\langle\mbox{Tr}\,\lambda^{\alpha}\lambda_{\alpha} \rangle /M_P^2$ in the case of a single confining 
group. By using the different gauge groups and matter contents which appear in string constructions it is 
possible to reach phenomenologically reasonable values of $m_{3/2}$, of order $1$ TeV \cite{deCarlosCM93}.

It has has usually been assumed that the structure of the gauge theory vacuum is well described by {\em 
globally}\/ supersymmetric QCD, since the strong-coupling scale is well below $M_P$. In the limit of 
global supersymmetry $M_P \rightarrow \infty$ the condensates do not break supersymmetry 
\cite{VenezianoY,TaylorVY,WittenIndex}. The condensate values may be found by instanton calculations 
\cite{ShifmanV+Amati88} or by constructing a supersymmetric effective action \cite{VenezianoY,TaylorVY} 
in terms of gauge invariant composite superfields $U\propto \mbox{Tr}\,W^\alpha W_\alpha$ and 
$V_{ij}\propto \bar{Q}_{ai} Q^a_j$, where $W^\alpha$ is the supersymmetric gauge field strength and 
$Q,\bar{Q}$ are the quark and antiquark chiral superfields respectively. The supersymmetric zero-energy 
vacuum satisfies the conditions $F_U = -\partial W^*_{np}/ \partial U^* =0$, $F_{V} = -\partial W^*_{np}/
 \partial V^*_{ij} =0$, where $W_{np}$ is the nonperturbative superpotential generated by the gauge 
dynamics. The gaugino condensate value is then determined (up to a discrete symmetry) in terms of the 
renormalisation group invariant scale $\Lambda$ and the squark mass matrix $M_{ij}$. In string-inspired 
supergravity models the scale $\Lambda$ and the squark masses are functions of the dilaton and moduli 
scalars of the underlying theory. Hence the supersymmetry-breaking has a non-trivial dependence on the 
``flat directions'' of string theory and it becomes possible to stabilize the moduli 
\cite{FontILQ90,FerraraMTV} and (under certain assumptions!) the dilaton 
\cite{BanksDine+Casas96+BGW+ChoiKK}, while inducing supersymmetry-breaking of the right magnitude 
\cite{deCarlosCM93,BrignoleIM+KaplL93}.  

Recently, SQCD has also been used in the construction of (supersymmetric) domain walls 
\cite{deC/MorenoDW}, motivated by cosmological issues or by the ``brane world'' scenario in which the 
observable matter fields are confined to a region localized in one or more directions within a 
higher-dimensional spacetime. The discrete set of degenerate vacua that is required is naturally realised 
in SU$(N)$ SQCD, since the vacuum with nonzero gluino and matter condensates transforms under a 
non-anomalous $\mathbb{Z}_N$ symmetry\footnote{This is the subgroup of the anomalous U$(1)_R$ that survives
breaking by instantons: in general there will be a $\mathbb{Z}_{c(G)}$ degeneracy.}.  When supersymmetry is 
broken by an explicit gluino mass term \cite{KovShifSmilga}, the degeneracy is also broken, which raises 
doubts as to the stability of domain walls when gravitational corrections to the scalar potential are 
included. A non-zero v.e.v.\ for $\mbox{Tr}\,\lambda^{\alpha}\lambda_{\alpha}$ will in general break (local)
SUSY and give the gluinos, including hidden sector gauginos, a mass. We will find, however, that 
the vacuum degeneracy is {\em unbroken}\/ when the source of SUSY-breaking is the gaugino condensate itself.

In this paper we attempt to find the corrections to the vacuum structure of SQCD due to its coupling to
$\mathcal{N}=1$ supergravity, including the dilaton and an overall modulus denoted by chiral
superfields $S$ and $T$ respectively. The starting point will be the non-perturbative superpotential
of \cite{TaylorVY} and the resulting zero-energy vacuum of the globally supersymmetric theory. It has
already been shown in the case of a SYM hidden sector {\em without}\/ matter, that the resulting value of
the gaugino condensate is identical to that at the minimum of the $\mathcal{N}=1$ supergravity scalar 
potential $\mathcal{V}(U,S,T)$, when terms in $U$ of mass dimension higher than $4$ ({\em i.e.\/}\ suppressed 
by powers of $M_P$) are discarded \cite{CasasLMR,deCarlosCM91}. We will show that this also holds in the 
case of SQCD: on discarding the terms in the supergravity potential
with mass dimension $>4$ in the condensate fields (note that we treat $S$ and $T$ as being
dimensionless), we recover the global SQCD vacuum. The size of deviations from the ``truncated
approximation'' for the condensates \cite{LustT91} can be found by minimising the scalar potential in 
supergravity as a function of $U$ and $V_{ij}$, including higher-order terms. A similar approach was taken 
in the case of pure SYM coupled to supergravity in \cite{FerraraMTV,LalakNN95,me99}.

\section{Effective action for the gaugino and squark condensates} 
We follow the approach of Burgess et al.\ \cite{BurgessDQQ} in which the gaugino condensate is described 
by the {\em classical}\/ field $U\equiv \hat{U}/S_0^3$, where $\hat{U} = \langle\mbox{Tr}\, 
W^{\alpha}W_{\alpha} \rangle$ and the chiral compensator superfield $S_0$ is introduced to simplify the 
formulation of supergravity coupled to matter \cite{SUGRA_matter} using the superconformal tensor 
calculus\footnote{The components of $S_0$ are determined by gauge-fixing the superconformal symmetries 
so that the Einstein term in the Lagrangian is canonically normalised \cite{BurgessDQQ}.}. The gaugino 
bilinear $\langle \mbox{Tr}\,\lambda^\alpha \lambda_\alpha \rangle$ corresponds to the lowest component 
of $\hat{U}$ ($\theta= \bar{\theta}=0)$. Similarly the squark condensate is the lowest component of a (mass 
dimension 2) composite superfield $V_{ij} = \langle \bar{Q}_{ai}Q^a_j \rangle$ where $Q_i$, 
$\bar{Q}_j$ are $N_f$ flavours of left chiral superfields in the representation $R$ and its complex 
conjugate $\bar{R}$ respectively. The v.e.v.'s of the gaugino and squark bilinears are then given by the 
scalar components of $U$ and $V_{ij}$ at the stationary point of the effective action $\Gamma(U,V,S,T)$ 
\cite{BurgessDQQ}. 

The effective action results from the standard $\mathcal{N}=1$ supergravity formula with the 
Taylor-Veneziano-Yankielowicz (TVY) nonperturbative superpotential in terms of $U$ and $V$, and an 
appropriate K\"ahler potential $K$. The TVY superpotential, suitably amended for the case of a 
non-minimal gauge kinetic function and a modulus-dependent mass matrix, is
\begin{equation}
	W(U,V_{ij},S,T)= \frac{1}{4}f_G(S,T) U - \frac{1}{96\pi^2}U \ln\left(kU^{b+2c}
	(\mbox{det}V_{ij})^{-3c/N_f}\right) - M_{ij}(T)V_{ij}. \label{eq:Wnp}
\end{equation}
Here $f_G(S,T)$ is the gauge kinetic function, equal to $S$ at tree level, which in general depends on 
the modulus $T$ through string loop threshold corrections \cite{DixonKL91}, $b$ is the one-loop beta 
function coefficient such that $b=-3c(G)+2N_f T(R)$, and $c=2N_f T(R)$; $c(G)$ is the second-order 
Casimir invariant for the gauge group $G$ and $T(R)$ is the index, equal to $1/2$ in the fundamental 
representation. $k$ is a constant which will be discussed shortly. It is convenient to diagonalise 
$V_{ij}$ by performing unitary flavour rotations of the matter fields $Q_i$, $\bar{Q}_j$ so that 
$V_{ij} = V_i \delta_{ij}$. The superpotential can then be re-expressed as  
\begin{equation}
	W = -\frac{b}{96\pi^2}U \ln\left(kU^{1+2c/b} \prod_iV_i^{-3c/N_f b} \omega(S)h(T)\right) - 
	M_i(T)V_i \label{eq:Wnp_h}
\end{equation}
where $\omega(S)\equiv e^{-24\pi^2S/b}$ and $h(T)$ is a function which transforms under the target-space 
duality group SL$(2,{\mathbb Z})$ \cite{FerraraMTV,Taylor90} such that the argument of the logarithm in 
(\ref{eq:Wnp_h}) is invariant. 

This is all we need to find the truncated approximation for the condensate values: it simply follows 
from setting the derivatives of $W$ with respect to $U$ and $V$ to zero, giving us 
\begin{equation}
	U^{(tr)} = e^{24\pi^2f_G(S,T)/b_0-(b+2c)/b_0} k^{-b/b_0} 
	\left(\frac{c}{32\pi^2N_f}\right)^{3c/b_0} \prod_iM_i(T)^{-3c/N_fb_0} 
	\label{eq:Utr}
\end{equation}
and
\begin{equation}
	V_i^{(tr)} = \frac{c}{32\pi^2N_f} \frac{U^{(tr)}}{M_i(T)} \label{eq:Vtr}	
\end{equation}
where $b_0=-3c(G)$ is the beta-function coefficient for SYM without matter. We see that the constant 
$k$ simply adjusts the scale of the condensates; it has the same effect as a constant threshold 
correction to $f_G$. 

When the condensate values (\ref{eq:Utr}) and (\ref{eq:Vtr}) are substituted back into the 
superpotential, the ``truncated superpotential'' $W_{np}^{(tr)}(S,T) = b_0/(96\pi^2)U^{(tr)}$ emerges. 
This is the usual starting point for studying supersymmetry-breaking in string effective field 
theories (see for example \cite{CveticFILQ}).

We take the K\"ahler potential for the dilaton and moduli to be
\begin{equation}
	\tilde{K} = P(y) -3 \ln(T+T^\dag) \label{eq:Ktilde0}
\end{equation}
where $y=S+S^\dag -3/(8\pi^2)\delta_{GS}\ln(T+T^\dag)$. The string tree-level K\"ahler potential for the 
dilaton $-\ln(S+S^\dag)$ has been replaced by a real function $P(y)$ which parameterizes stringy 
nonperturbative dilaton dynamics \cite{Shenker90,BanksDine+Casas96+BGW+ChoiKK}. We take the 
Green-Schwarz coefficients $\delta_{GS}$ to be zero to simplify the calculations. The correct form of 
$P(y)$ is not known, however it is possible to constrain it by looking for a stable minimum in the 
potential for the dilaton and requiring $P''(y)>0$ to obtain the right sign kinetic term.

The $\mathcal{N}=1$ supergravity effective action is invariant under target-space SL$(2,{\mathbb Z})$ 
transformations \cite{FerraraLST,FerraraLT_2} if the superpotential transforms as a modular 
form\footnote{For a discussion of modular forms see \cite{Rankin} or the appendix of 
\cite{CveticFILQ}.} of weight $-3$. The modulus $T$ transforms as 
\begin{equation}
	T \rightarrow \frac{\alpha T-i\beta}{i\gamma T+\delta} \nonumber
\end{equation}
where $\alpha,\beta,\gamma,\delta$ are integers satisfying $\alpha\delta-\beta\gamma = 1$. Then $U$ 
must transform as
\begin{equation} 
	U \rightarrow \zeta(i\gamma T+\delta)^{-3} U, \label{eq:Utransf} 
\end{equation} 
where $\zeta$ is a unimodular phase which depends on $\alpha,\beta,\gamma$ and $\delta$. Since the gauge 
fields are inert under target-space modular transformations this is achieved by $S_0$ having a 
non-trivial transformation property. The transformation of the squark condensate $V$ is determined by 
that of the $Q$, $\bar{Q}$ fields, which we take to have (flavour-independent) modular weights 
$n_Q$, $n_{\bar{Q}}$: then $V_i \rightarrow \zeta_V(i\gamma T+\delta)^{n_Q+n_{\bar{Q}}} V_i$. The function 
$h(T)$ is determined by string threshold corrections (up to multiplication by a modular invariant 
function of $T$) as $h(T) = \eta(T)^{6b'/b}$, where $\eta(T)$ is the Dedekind eta-function and the 
coefficient $b'$ is \cite{LustMunoz92} $b' = 3(-c(G) + \sum_Q T(R_Q)(1+2n_Q))$ for a gauge group with 
twisted matter representations $R_Q$ of modular weight $n_Q$. 

The complete K\"ahler potential is then taken to be 
\begin{equation}
	K(U,V_i,S,T) = \tilde{K} - 3\ln\left(1- A e^{\tilde{K}/3} (UU^\dag)^{1/3}\right) + 
B(T+T^\dag)^{\bar{n}} (V_iV_i^\dag)^{1/2}. \label{eq:K1} 
\end{equation}  
where $A$ and $B$ are constants and $\bar{n}\equiv(n_Q+n_{\bar{Q}})/2$. It was shown in 
\cite{BurgessDQQ} that this expression for the K\"ahler potential of $U$ has the correct dependence on 
$S$ and $T$; the $V$-dependent part is fixed by the requirements to respect U$(N_f)$ flavour symmetry 
and modular invariance. The K\"ahler potential for the composite fields $U$ and $V$ is only determined 
up to constant factors, and may receive higher-order corrections (which, however, will be negligible 
when the field values are small). The constants $A$, $B$ cannot at present be computed, due to our 
incomplete knowledge of supersymmetric gauge dynamics, but they are expected to be of order unity. 
Since the scalar potential in supergravity is a function of $K$ as well as the superpotential $W$ these 
constants will also appear in our results. In an earlier paper \cite{me99} the constant $A$ was 
absorbed by rescaling the gaugino condensate, however since physical SUSY-breaking quantities are 
expressed in terms of the {\em un-rescaled}\/ gaugino condensate (and, in general, the normalisation of 
fields is fixed by the coefficients with which they enter into $W$) this will not be done here.

The part of $\Gamma(U,V_i,S,T)$ relevant to finding the value of the condensate is the scalar 
potential, which is given as usual by\footnote{We work in reduced Planck units with $\kappa^{-1} 
=1/\sqrt{8\pi G}$ set to $1$.}
	\[ \mathcal{V} = e^K\left((W^*_I+K_IW^*) (K^{-1})^I_J (W^J+K^JW) - 3|W|^2\right) \]
where $I$ and $J$ range over the scalar components of $U$, $V_i$, $S$ and $T$, $X^I\equiv \partial 
X/\partial \phi_I$ and $X_J\equiv \partial X/\partial\phi^{J*}$ for $X$ any function of the scalars and 
their complex conjugates, and $K^{-1}$ is defined by $(K^{-1})^I_J K^J_K = \delta^I_K$. The details of 
the inverse K\"ahler metric components are relegated to the Appendix.

As emphasized in \cite{Kov/Shifman}, $\Gamma(U,V_i,S,T)$ is not an effective Lagrangian in the sense of 
describing the light degrees of freedom only, (it is a functional of the heavy fields $U$ and 
$V$), rather it is the generating functional of ``two-particle irreducible'' correlation functions 
for the composite fields $\mbox{Tr}\,W^2$ and $\mbox{Tr}\,\bar{Q_i}Q_j$ (see also \cite{BurgessDQQ}). The 
kinetic terms in $\Gamma$ must be understood as the first terms in a derivative expansion; there may be
large higher-derivative corrections, so the effective action cannot be reliably used to determine 
particle interaction vertices. However the scalar potential, {\em i.e.\/}\ $\Gamma (U,V_i,S,T)$ for constant 
field configurations, does not get these corrections. 

After some calculation, we find the scalar potential in closed form, which for convenience is expressed 
in terms of the rescaled quantities $z= e^{\tilde{K}/6}U^{1/3}$, $\Pi_i=(T+T^*)^{\bar{n}}V_i$, 
$\|\Pi\|=(\Pi_i\Pi_i^*)^{1/2}$:
\begin{equation} 
	\mathcal{V} = \frac{e^{B\|\Pi\|}}{(1-A|z|^2)^3} \left\{\mathcal{V}_0 + \mathcal{V}_1\right\} 
\label{eq:V01}
\end{equation}
with
\begin{eqnarray}
	\mathcal{V}_0 &=& \left(\frac{b}{96\pi^2}\right)^2 \frac{3|z|^4}{A} (1-A|z|^2) \left| 
1+\frac{2c}{b} + \mathcal{LOG}\right|^2 
\nonumber \\
	&+& \frac{2}{B\|\Pi\|} \left(\delta_{ij}\|\Pi\|^2+\Pi^*_i\Pi_j\right) 
\left(\frac{c}{32\pi^2N_f} z^{*3}\Pi_i^{*-1} - e^{\tilde{K}/2}(T+T^*)^{-\bar{n}} M_i(T^*)\right)  
\nonumber \\
	& & \times \left(\frac{c}{32\pi^2N_f} z^{3}\Pi_j^{-1} - e^{\tilde{K}/2}(T+T^*)^{-\bar{n}} 
M_j(T)\right), \label{eq:V0}
\end{eqnarray}
\begin{eqnarray}
	\mathcal{V}_1 &=& \frac{P'(y)^2}{P''(y)}(1-A|z|^2) \mathcal{V}_S + 
\frac{1-A|z|^2}{3(1-\frac{\bar{n}}{3}(1-A|z|^2)B\|\Pi\|)} 
\mathcal{V}_T 
\nonumber \\
	&+& \left(\frac{b}{96\pi^2}\right)^2 |z|^6 \left(-3\left(1+\frac{2c}{b}\right)^2 
-\frac{6c}{b}(\mathcal{LOG}+c.c.) + B\|\Pi\|\cdot|\mathcal{LOG}|^2 \right) \nonumber \\
	&+& \frac{b}{96\pi^2} \left(2+B\|\Pi\|\right) \left(e^{\tilde{K}/2}(T+T^*)^{-\bar{n}} z^3 
	\mathcal{LOG} \cdot M_i(T^*)\Pi^*_i + c.c.\right) \nonumber \\ 
	&+& \frac{3b}{96\pi^2} \left(1-A|z|^2\left(1+\frac{2c}{b}\right)\right) 
\left(e^{\tilde{K}/2}(T+T^*)^{-\bar{n}} z^3 M_i(T^*)\Pi^*_i + c.c.\right) \nonumber \\
	&+& e^{\tilde{K}}(T+T^*)^{-2\bar{n}} \left(1+3A|z|^2+B\|\Pi\|\right) \left|M_i(T)\Pi_i\right|^2 
	\label{eq:V1}
\end{eqnarray}
where
\begin{eqnarray}
	\mathcal{LOG} &=& \ln(kU^{1+2c/b} \prod_iV_i^{-3c/N_f b} \omega(S)h(T)), \label{eq:Log} \\
	\mathcal{V}_S &=& \left|\frac{b}{96\pi^2}z^3 \left(\frac{1}{P'(y)} \frac{\omega'(S)}{\omega(S)} 
-\left(1+\frac{2c}{b}\right)\right) + e^{\tilde{K}/2}(T+T^*)^{-\bar{n}} M_i(T)\Pi_i\right|^2, \nonumber 
\\
	\mathcal{V}_T &=& \left|\frac{b}{96\pi^2}z^3 \left((T+T^*)\frac{h'(T)}{h(T)} + 
3\left(1+\frac{2c}{b}(1+\bar{n})\right)\right) \right. + \nonumber \\
	&+& \left. e^{\tilde{K}/2}(T+T^*)^{-\bar{n}} \left((T+T^*)\frac{M'_i(T)}{M_i(T)} 
-(3+2\bar{n})\right) M_i(T)\Pi_i \right|^2.
\end{eqnarray}
In this somewhat unwieldy expression, $\mathcal{V}_0$ includes the terms of mass dimension 4 [glossing 
over factors of $(1-A|z|^2)$], which survive in the global supersymmetry limit $M_P \rightarrow 
\infty$, $z \rightarrow 0$, $\Pi \rightarrow 0$, while $\mathcal{V}_1$ includes all terms of higher 
order in $z$ and $\Pi$. Note also the ``modular covariant'' expressions 
	\[ \hat{G}^T\equiv (T+T^*)\frac{h'(T)}{h(T)} +3\left(1+ \frac{2c}{b}(1+\bar{n})\right)\mbox{, } 
\hat{G}^{M_i}\equiv (T+T^*)\frac{M'_i(T)}{M_i(T)} -(3+2\bar{n}) \]
appearing in $\mathcal{V}_T$, which transform into themselves up to a phase under SL$(2,{\mathbb Z})$. For 
future convenience we also define 
	\[ \hat{G}^S\equiv \frac{1}{P'(y)} \frac{\omega'(S)}{\omega(S)} -\left(1+\frac{2c}{b}\right). \]

As expected, $\mathcal{V}_0$ has a zero-value minimum in the condensate directions\footnote{Note that 
the {\em principal value}\/ of the logarithm in (\ref{eq:Log}) must be taken. Since the argument is real 
{\em in vacuo}\/ this does not cause an ambiguity.} at $z=z^{(tr)}\equiv e^{\tilde{K}/6}U^{(tr)1/3}$ and 
$\Pi_i=\Pi_i^{(tr)}\equiv (T+T^*)^{\bar{n}} V_i^{(tr)}$. The ``truncated superpotential'' result for 
the scalar potential is then equivalent to substituting these condensate values back into the full 
potential, which is then proportional to $\mathcal{V}_1$, and discarding terms of dimension greater 
than 6 in powers of $z$ and $\Pi_i$. It is a non-trivial check on the result (\ref{eq:V1}) to carry out this 
substitution and verify that the scalar potential then coincides with previous results 
\cite{Taylor90,deCarlosCM91}. In fact we obtain
\begin{eqnarray*} 
	\mathcal{V}^{(tr)} &\equiv& \mathcal{V}(S,T)_{|z=z^{(tr)},\Pi=\Pi^{(tr)}} = 
	\left(\frac{b_0}{96\pi^2}\right)^2 |z^{(tr)}|^6 \left\{ \frac{1}{P''(y)} 
	\left|\frac{\omega'_0(S)}{\omega_0(S)} -P'(y)\right|^2 + \right. \\
	&+& \left. \frac{1}{3}\left|(T+T^*)\frac{h'_0(T)}{h_0(T)} + 3\right|^2 - 3 \right\} + 
	\mathcal{O}(z^{(tr)8}) 
\end{eqnarray*}
where $\omega_0(S) = e^{-24\pi^2S/b_0}$, $h_0(T) = \eta(T)^{6b'/b_0}\prod_iM_i(T)^{3c/N_fb_0}$. Note 
that the unknown constants $A$ and $B$ drop out of this expression. This occurs because the truncated 
approximation for the condensates is implemented via the superpotential $W_{np}$ alone, and does not 
(at the given order of approximation in $z/M_{P}$) affect the K\"ahler potential for $S$ and $T$.

\section{Beyond the truncated approximation}
The truncated approximation relies on the assumption that the condensate values at the minimum of 
$\mathcal{V}_0$ is not much changed when the gravitationally-suppressed terms in $\mathcal{V}_1$ are 
added. This is likely to be true when the condensate values $z$, $\Pi$ are small, which is also the 
phenomenologically interesting regime for most applications. We are able to test this assumption 
explicitly by expanding about the truncated values to find the shift in the minimum. As a simplifying 
assumption the masses $M_i(T)$ are set equal to $M(T)$ and the squark condensates $\Pi_i$ equal to 
$\Pi$. We define the fractional changes in the condensate values 
	\[ \Delta z = \frac{z}{z^{(tr)}}-1,\qquad \Delta \Pi = \frac{\Pi}{\Pi^{(tr)}}-1 \]
and take the real and imaginary parts Re$\,\Delta z$, Im$\,\Delta z$, Re$\,\Delta\Pi$, Im$\,\Delta\Pi$ as
independent variables. We then 
expand the scalar potential (\ref{eq:V01}) up to terms of mass dimension $6$ in the fields (in fact, we 
would expect higher-order corrections to the K\"ahler potential for $U$ and $V_i$ to become significant 
beyond this order) and up to quadratic order in $\Delta z$ and $\Delta\Pi$. The result takes the form
\begin{equation} 
	\mathcal{V}= \mathcal{V}^{(tr)} + \vec{\mathcal{V}}_L^T \cdot\vec{\Delta} + \vec{\Delta}^T 
\underline{\underline{\mathcal{V}_Q}} 
\vec{\Delta}
\end{equation}
where $\vec{\Delta}$ is the column-vector with entries Re$\,\Delta z$, Im$\,\Delta z$, Re$\,\Delta\Pi$, 
Im$\,\Delta\Pi$, and $\vec{\mathcal{V}}_L$ and $\underline{\underline{\mathcal{V}_Q}}$ are a vector and 
symmetric matrix of coefficients respectively. The displacements $\vec{\Delta}_{min}$ at the minimum of 
$\mathcal{V}$ are given by
\begin{equation} 
	\vec{\Delta}_{min} = -\frac{1}{2} \underline{\underline{\mathcal{V}_Q}}^{-1} \cdot 
\vec{\mathcal{V}}_L
\end{equation}
and the change in the potential from the truncated value is 
\begin{equation}
	\mathcal{V}_{min} - \mathcal{V}^{(tr)} = -\frac{1}{4} \vec{\mathcal{V}}_L^T 
\underline{\underline{\mathcal{V}_Q}}^{-1} \vec{\mathcal{V}}_L.
\end{equation}

The coefficients in $\underline{\underline{\mathcal{V}_Q}}$ are of dimension four and six in the fields 
(and in the modulus-dependent mass $M(T)$) while those in $\vec{\mathcal{V}}_L$ are dimension six only 
(since they originate from $\mathcal{V}_1$). Since we are evaluating $\vec{\Delta}_{min}$ and 
$\mathcal{V}_{min}$ to leading order in $z^{(tr)}$ we can discard dimension six terms in 
$\underline{\underline{\mathcal{V}_Q}}$, provided that the matrix does not become singular. Then we 
have
\begin{eqnarray}
	\mathcal{V}_L^T \cdot\vec{\Delta} &=& \frac{|z^{(tr)}|^6}{(32\pi^2)^2} \left[ \mbox{Re}\,\Delta 
z\left(\frac{P'(y)^2}{P''(y)}\left(\frac{2b^2}{3} ({\hat{G}^S})^2 + 2bc\hat{G}^S\right) + 
\frac{2b^2}{9}|\hat{G}^T|^2 + \right.\right. \nonumber \\
	&+& \left.\left. \frac{2bc}{3} \mbox{Re}\,(\hat{G}^T\hat{G}^{M*}) -2b_0 b  \right) 
+ \mbox{Im}\,\Delta z \cdot -\frac{2bc}{3}\mbox{Im}\,(\hat{G}^T\hat{G}^{M*}) \right. \nonumber \\
	&+& \left. \mbox{Re}\,\Delta\Pi \left(\frac{P'(y)^2}{P''(y)}\left(\frac{2bc}{3}\hat{G}^S 
	+2c^2 \right) + \frac{2c^2}{3}|\hat{G}^M|^2 +  \right.\right. \nonumber \\
	&+& \left.\left. \frac{2bc}{9}\mbox{Re}\,(\hat{G}^T\hat{G}^{M*}) + \frac{2b_0c}{3} \right) 
	+\mbox{Im}\,\Delta\Pi \cdot \frac{2bc}{9} \mbox{Im}\,(\hat{G}^T\hat{G}^{M*}) \right]
\end{eqnarray}
and
\begin{eqnarray}
	\vec{\Delta}^T \underline{\underline{\mathcal{V}_Q}} \vec{\Delta} &=& 
\frac{|z^{(tr)}|^4}{(32\pi^2)^2} \left[ \frac{3}{A} \left( (b+2c)^2 ((\mbox{Re}\,\Delta z)^2+ (\mbox{Im}\,\Delta z)^2) 
\right.\right. \nonumber \\
	&-&  \left. 2c(b+2c) (\mbox{Re}\,\Delta z \mbox{Re}\,\Delta\Pi + \mbox{Im}\,\Delta z\mbox{Im}\,\Delta\Pi) + c^2 
((\mbox{Re}\,\Delta\Pi)^2+ (\mbox{Im}\,\Delta\Pi)^2) \right) \nonumber \\
	&+&  \frac{4(32\pi^2)cN_f^{1/2}e^{\tilde{K}/2}|M(T)|}{B|z^{(tr)}|(T+T^*)^{\bar{n}}} 
\left(9((\mbox{Re}\,\Delta z)^2+ (\mbox{Im}\,\Delta z)^2) +  \right. \nonumber \\
	&+& \left.\left. (\mbox{Re}\,\Delta\Pi)^2+(\mbox{Im}\,\Delta\Pi)^2 \right) + \mathcal{O}(|z^{(tr)}|^6) 
	\rule{0cm}{3ex} \right]
\end{eqnarray}
Note that the quadratic terms originating from $\mathcal{V}_0$ fall into two parts, reflecting their 
origins in $|F_U|^2$ and $|F_V|^2$, and there is an explicit dependence on the mass $M(T)$. Without the 
terms proportional to $M$, the matrix $\underline{\underline{\mathcal{V}_Q}}$ would be singular, 
reflecting the runaway instability in {\em massless}\/ globally supersymmetric QCD. With the mass terms 
$\underline{\underline{\mathcal{V}_Q}}$ can easily be inverted and it is a matter of algebra to find 
$\vec{\Delta}_{min}$ and $\mathcal{V}_{min}$. 

If the functions $G^T$ and $G^M$ have the same phase then the deviations $\Delta z$ and $\Delta\Pi$ are 
real. In this case we can get an intuitive idea by plotting 
\begin{figure}
\centering
\includegraphics[width=10.5cm,height=8cm]{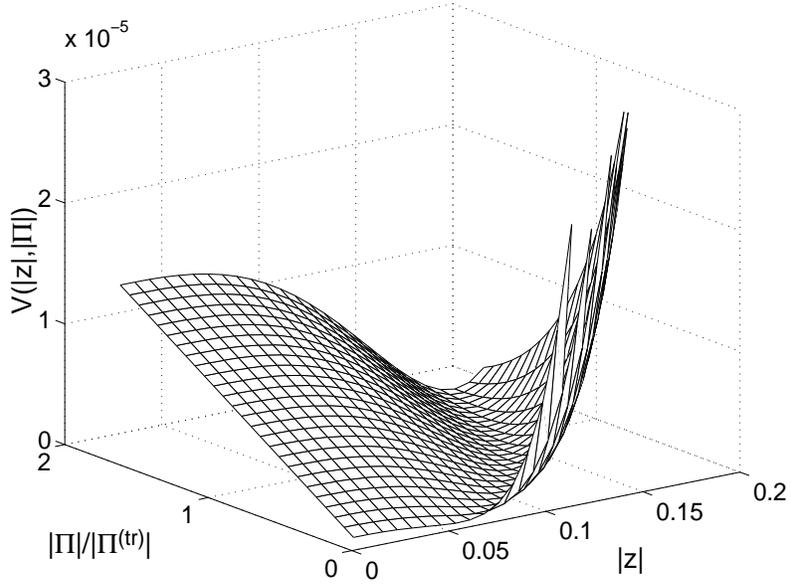}
\caption{Full scalar potential $\mathcal{V}(|z|,|\Pi|)$ ($|z^{(tr)}|=0.15,|M|=0.07$, other parameters as
 specified in the text)}
\end{figure}
the dependence of the full potential (\ref{eq:V01}) on the absolute values of the condensates $|z|,|\Pi|$ 
(given that their phases are fixed to the truncated values). Figure 1 shows the scalar potential 
$\mathcal{V}(|z|,|\Pi|)$ for $b_0 = -12$, $c=2$, $|z^{(tr)}|=0.15$ and $M=0.07$, with $\hat{G}^S$, 
$\hat{G}^T$, $\hat{G}^M$ set to the (somewhat arbitrary) values of $0.5$, $3.0$, $0.0$ respectively. Figure 2 
\begin{figure}
\centering
\includegraphics[width=10.5cm,height=8cm]{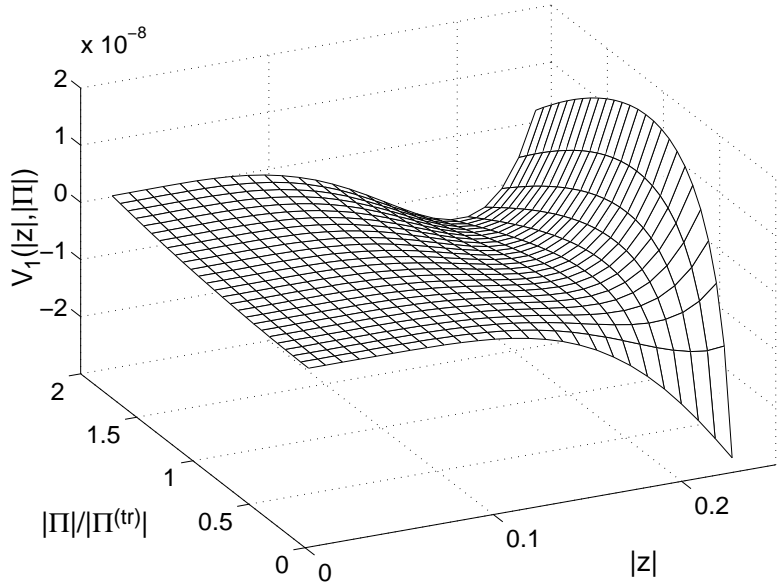}
\caption{Supergravity corrections $\mathcal{V}_1(|z|,|\Pi|)$}
\end{figure}
shows the gravitationally-suppressed corrections $\mathcal{V}_1(|z|,|\Pi|)$ for the same values of 
parameters. The structure of these terms is much richer than in the case without matter multiplets: 
however along the curve specified by $\partial W/\partial\Pi=0$, the form of the scalar potential and 
supergravity corrections 
\begin{figure}
\centering
\includegraphics[width=8.5cm,height=6.5cm]{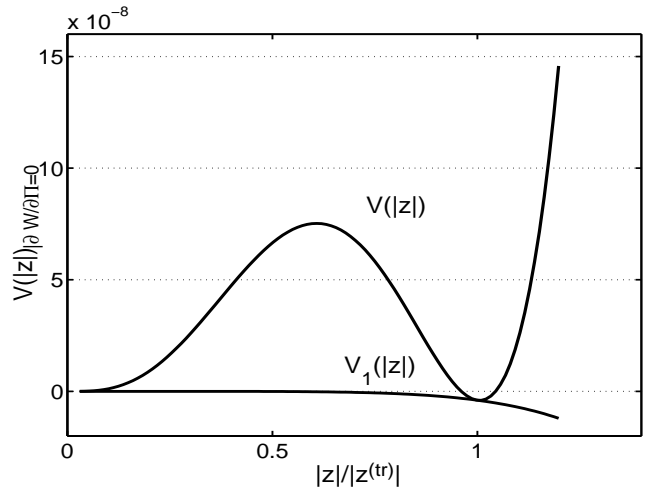}
\caption{$\mathcal{V}(|z|)$ and $\mathcal{V}_1(|z|)$ along the curve $\partial W/\partial\Pi=0$ for 
$|z|^{(tr)}=0.15$ and a negative cosmological constant.}
\end{figure}
reduces to the case of pure SYM (fig.\ 3) (see {\em e.g.}\/\ \cite{Taylor90}) .


There remains the question of the fate of the $N$ degenerate vacua in SU$(N)$ SQCD when supersymmetry 
is broken. Under the $\mathbb{Z}_N$ symmetry the condensates transform as $U \rightarrow e^{2i\pi/N} U$, 
$V_i \rightarrow e^{2i\pi/N} V_i$. It is known \cite{Veneziano95,Kov/Shifman} that the logarithmic term 
in the TVY effective action does not respect the symmetry, since the branch cut is crossed. There are 
in practice various ways of restoring the symmetry for the $\mathbb{Z}_N$ transformed vacua, including 
giving the squark condensates extra $2\pi$ phases (see \cite[first reference]{deC/MorenoDW}) such 
that the argument of the logarithm is invariant, adding extra fields \cite{Kov/Shifman}, or shifting the 
theta-angle by $2\pi$ (equivalent to taking $S \rightarrow S+i/8\pi^2$) which will exactly cancel the 
effect of an axial $\mathbb{Z}_N$ transformation. Taking such ``fixes'' into account, the scalar potential 
(\ref{eq:V01}) is invariant under the $\mathbb{Z}_N$ symmetry by inspection, since $z^3$ and $\Pi_i$ get the 
same phase. While this is a reassuring result for the construction of domain walls, we must ask why it 
appears to contradict the reasoning that any gaugino mass must break the vacuum degeneracy. We certainly 
expect that the ({\em hidden sector}\/) gauginos will get a small (in general complex) SUSY-breaking mass 
in the case we have studied. However, since the source of the mass is the gaugino condensate itself, {\em 
the gaugino mass is not inert under the}\/ $\mathbb{Z}_N$ {\em symmetry}\/. In fact the mass term 
will be proportional to $z^{*3}\mbox{Tr}\,\lambda\lambda$ (see for example 
\cite[first reference]{BrignoleIM+KaplL93}), which {\em is}\/ invariant\footnote{To see this, it is necessary 
to use the formulas for the $\mathcal{N}=1$ supergravity action in terms of $W$ and $K$ separately, rather 
than the K\"ahler function $\mathcal{G}=K+\log|W|^2$, since in going from one formulation to the other the 
gaugino fields are redefined by a complex phase and the phase of $W$ is eliminated \cite{SUGRA_SYM}.}.

\section{Conclusions}
Given an explicit form for the K\"ahler potential of the composite superfields representing the 
condensates, the scalar potential in a string-inspired ${\mathcal N}=1$ supergravity model can be 
calculated; this allows us to find the minimum in the condensate directions without any assumptions 
about the presence or absence of supersymmetry-breaking. The deviation from the ``truncated'' globally 
supersymmetric vacuum of SQCD have been found in terms of the dilaton and string moduli. The 
phenomenological implications of deviations from the truncated approximation were discussed in an 
earlier paper \cite{me99}. In the case with hidden matter, there is another possible cause for $\Delta z$, 
$\Delta\Pi$ to be large, namely if $\hat{G}^M$ becomes large due to singular behaviour of the mass 
$M(T)$ at some points in moduli space. Since $\det \underline{\underline{\mathcal{V}_Q}}$ is proportional 
to $M$, the deviations may also be large if $M$ becomes small. However, in the limit $M \ll z$ the 
SQCD vacuum undergoes a phase transition and the charged matter fields themselves (rather than the composite 
operator $\mbox{Tr}\,\bar{Q}Q$) can get v.e.v's, so our effective action will not be valid.

While the particular model from which this potential is derived, inspired by orbifolds of the heterotic 
string, is quite restrictive in its dependence on the dilaton and moduli, it would be relatively simple 
to extend the result to the case when the gauge kinetic function and mass matrix depend on several 
moduli, or the K\"ahler potential for the moduli is modified from its tree-level form ({\em e.g.\/}\ by 
introducing functions $P_i(T_i+T_i^\dag)$, where $T_i$ is the $i$'th modulus). So our approach should be
applicable to supergravity effective theories based on heterotic M-theory \cite{HW+Munoz99} or Type 
IIB string models \cite[and references therein]{Ibanez99}. In particular we expect that the supergravity 
corrections will preserve the discrete symmetry of degenerate vacua in SYM and SQCD.

\section*{Acknowledgements}
Thanks are due to David Bailin for supervising the work and Beatriz de Carlos for an enlightening 
discussion. TD is supported by PPARC studentship PPA/S/S/1997/02555.

{\appendix
\section*{Appendix: Inverse K\"ahler metric components}
The inverse K\"ahler metric components are as follows:
\[ (K^{-1})^U_U = \frac{|z|^4(1-A|z|^2)}{Ae^{\tilde{K}}} 
	\left(3+\frac{\bar{n}AB|z|^2(1-A|z|^2)\|\Pi\|}
	{1-\frac{\bar{n}}{3}B(1-A|z|^2)\|\Pi\|} + \frac{A|z|^2P'(y)^2}{P''(y)} \right), \]
\[ (K^{-1})^U_{V_i} = \frac{-2\bar{n}z^{*3}(1-A|z|^2)\Pi_i} 
	{e^{\tilde{K}/2}(T+T^*)^{\bar{n}}(1-\frac{\bar{n}}{3}B(1-A|z|^2)\|\Pi\|)}, \]
\[ (K^{-1})^U_S = -\frac{z^{*3}(1-A|z|^2)P'(y)} {e^{\tilde{K}/2}P''(y)},\mbox{ }
	(K^{-1})^U_T = \frac{(T+T^*)z^{*3}(1-A|z|^2)} {e^{\tilde{K}/2} 
	(1-\frac{\bar{n}}{3}B(1-A|z|^2)\|\Pi\|)}, \]
\[ (K^{-1})^{V_i}_{V_j} = \frac{2}{B(T+T^*)^{2\bar{n}}\|\Pi\|} \left(\delta_{ij}\|\Pi\|^2 + 
	\Pi^*_i\Pi_j \frac{1-\frac{\bar{n}}{3}(1-2\bar{n})(1-A|z|^2)B\|\Pi\|} 
	{1-\frac{\bar{n}}{3}(1-A|z|^2)B\|\Pi\|} \right), \]
\[ (K^{-1})^{V_i}_S = 0 = (K^{-1})^S_T,\mbox{ }
   (K^{-1})^{V_i}_T = -\frac{2\bar{n}(T+T^*)(1-A|z|^2)\Pi^*_i}
	{3(T+T^*)^{\bar{n}} (1-\frac{\bar{n}}{3}B(1-A|z|^2)\|\Pi\|)},\]
\[ (K^{-1})^S_S = \frac{1-A|z|^2}{P''(y)},\qquad
   (K^{-1})^T_T = \frac{(T+T^*)^2(1-A|z|^2)} {3(1-\frac{\bar{n}}{3}B(1-A|z|^2)\|\Pi\|)} \]
where $z= e^{\tilde{K}/6}U^{1/3}$, $\Pi_i=(T+T^*)^{\bar{n}}V_i$, $\|\Pi\|=(\Pi_i\Pi_i^*)^{1/2}$.
}

\end{document}